\documentclass[twocolumn,floatfix,prb,aps,showpacs]{revtex4-2}
\usepackage{graphicx,amsmath,amssymb,color,nicefrac,multirow, makecell, ragged2e,float}
\usepackage[unicode=true,colorlinks=true,linkcolor=blue,citecolor=blue,urlcolor=blue]{hyperref}
\usepackage{bm,physics,enumerate,booktabs,xcolor,float,ulem}
\usepackage[titletoc,title]{appendix}

\newcommand{\be}{\begin{equation}}
\newcommand{\ee}{\end{equation}}

\newcommand{\ba}{\begin{eqnarray}}
\newcommand{\ea}{\end{eqnarray}}

\renewcommand{\vec}[1]{\mbox{\boldmath$#1$}}

\def\beq{\begin{eqnarray}}
\def\eeq{\end{eqnarray}}

\makeatletter
\newcommand*{\rom}[1]{\expandafter\@slowromancap\romannumeral #1@}
\makeatother

\newcommand{\non}{\nonumber\\}

\newcommand{\ave}[1]{\langle{#1}\rangle}
\allowdisplaybreaks[4]

\begin{document}
\title{Candidate local parent Hamiltonian for $3/7$ fractional quantum Hall effect}
\author{Koji Kudo$^{1,2}$, A. Sharma$^1$, G. J. Sreejith$^3$ and J. K. Jain$^1$}
\affiliation{$^1$Department of Physics, The Pennsylvania State University, University Park, Pennsylvania 16802, USA}
\affiliation{$^2$Department of Physics, Kyushu University, Fukuoka 819-0395, Japan}
\affiliation{$^3$Indian Institute of Science Education and Research, Pune 411008, India}
\begin{abstract}
While a parent Hamiltonian for the Laughlin $1/3$ wave function has been long known in terms of the Haldane pseudopotentials, no parent Hamiltonians are known for the lowest-Landau-level projected wave functions of the composite fermion theory at $n/(2n+1)$ with $n\geq2$.  If one takes the two lowest Landau levels to be degenerate, the Trugman-Kivelson interaction produces the {\it unprojected} 2/5 wave function as the unique zero energy solution.  If the lowest three Landau levels are assumed to be degenerate, the Trugman-Kivelson interaction produces a large number of zero energy states at $\nu=3/7$. 
We propose that adding an appropriately constructed three-body interaction yields the unprojected $3/7$ wave function as the {\it unique} zero energy solution, and report extensive exact diagonalization studies that provide strong support to this proposal. 
\end{abstract}
\maketitle
\section{Introduction}

The fractional quantum Hall effect (FQHE)~\cite{Tsui82} is one of the most striking phenomena to arise from the interaction between electrons. 
Its rich phenomenology is explained in terms of emergent particles called composite fermions (CFs), which are bound states of electrons and an even number of quantized vortices, sometimes viewed as electrons bound to an even number of magnetic flux quanta~\cite{Jain89}. A remarkable aspect of the CF theory is that it establishes a mapping between the FQHE of electrons at filling factors $\nu=n/(2pn\pm 1)$ and the integer quantum Hall effect (IQHE) of CFs carrying $2p$ vortices at filling factor $\nu^*=n$. 
It further allows explicit construction of wave functions for the FQHE states starting from the known IQHE wave functions~\cite{Jain89}, which provide extremely accurate representations of the exact Coulomb wave functions known numerically for finite systems for which exact diagonalization on the computer is possible~\cite{Jain07,Halperin20}. 
The Laughlin wave function~\cite{Laughlin83} appears in this theory as the ground state wave function of CFs at filling factor $\nu^*=1$.

While the close agreement with the Coulomb solutions is sufficient to establish the quantitative validity of these wave functions, one may ask if they are exact solutions of some model interactions. The interaction is often expressed in terms of the Haldane pseudopotentials $V_m$, which are energies of pairs of electrons with relative angular momenta $m$~\cite{Haldane83}. Haldane showed that the Laughlin $1/3$ state is the exact and unique zero energy state of fully spin polarized electrons confined to the lowest Landau level (LLL) for the interaction $V_m=\delta_{m,1}$~\cite{Haldane83}. 

The Jain wave functions at $\nu=n/(2pn\pm 1)$ are given by 
\begin{equation}
\Psi^{\rm LLL}_{n/(2pn\pm1)}
 ={\cal P}_{\rm LLL}\Phi_{\pm n}\Phi_1^{2p},
\end{equation}
where $\Phi_{+n}$ is a Slater determinant state 
of completely filled lowest $n$ LLs, we define $\Phi_{-n}=\Phi_{-n}^*$,
and ${\cal P}_{\rm LLL}$ denotes projection into the LLL. 
(We drop the ubiquitous Gaussian factors, which will be absent anyway once we specialize to the spherical geometry.)
Extensive studies~\cite{Sreejith18} have failed to find a pseudopotential Hamiltonian for which the LLL projected Jain 2/5 wave function for fermions, or the analogous 2/3 wave function for bosons, is the exact ground state.  

We will consider below the {\it unprojected} Jain $n/(2pn+ 1)$ wave functions, 
 referred to simply as the $n/(2pn+1)$ wave functions below,
given by 
\begin{equation}
 \Psi^{\rm }_{n/(2pn+1)} =\Phi_{n}\Phi_1^{2p}.
\end{equation}
These have a simpler form, but involve higher Landau levels (LLs). 
(The number of LLs participating in a wave function can be read off from the highest power of $\bar{z}_j$ in the polynomial part of the wave function multiplying the Gaussian factor; the highest power $\bar{z}^m$ implies nonzero weight in the lowest $m+1$ LLs.)
Interestingly, the Trugman-Kivelson (TK) interaction~\cite{Trugman85} 
\begin{align}
 V_\text{TK}=\nabla^2_2\delta^{(2)}(\vec{r}_2-\vec{r}_1)
 \label{eq:TK}
\end{align}
obtains the 2/5 wave function as the exact and unique zero energy ground state provided that the lowest two LLs are taken to be degenerate~\cite{Jain90,Rezayi91}.
(Here, $\nabla^2_2$ represents the 
Laplacian with respect to $\vec{r}_2$.)
One can see, using integration by parts, that any state that vanishes as 
the first power of the distance
$r$ between two particles, when they are brought close to one another, has a finite energy for $V_\text{TK}$, but any state that vanishes as $r^3$ has zero energy. 
(States that vanish as $r^2$ are not allowed due to antisymmetry of the fermionic wavefunctions.)
One can further show that $\Phi_2\Phi_1^2$ is the only state at $\nu=2/5$ 
that vanishes as $r^3$ within the space of the lowest two LLs . Numerical diagonalization has shown that this state evolves smoothly, without gap closing, for either the short-range or the Coulomb interaction as the kinetic energy gap between the lowest two LLs is increased from zero to infinity~\cite{Rezayi91}.

This strategy does not carry over to the 3/7. Because the unprojected Jain wave function at $\nu=3/7$ involves the lowest three LLs, we assume the lowest three LLs to be degenerate. The 3/7 wave function, $\Psi_{3/7}=\Phi_3\Phi_1^2$,
vanishes as $r^3$ (as each factor vanishes as $r$) when two particles approach one another and thus
has zero energy for the TK interaction, but it is not the only wave function with this property. It is degenerate with many other states of the form:
\begin{align}
 \Psi^{(\nu^*_1,\nu^*_2)}_{3/7}=\chi_{\nu^*_1}\chi_{\nu^*_2}\Phi_1,
 \label{eq:nunu}
\end{align}
where $\nu^*_1$ and $\nu^*_2$ satisfy $(\nu^*_1)^{-1}+(\nu^*_2)^{-1}+1^{-1}=(3/7)^{-1}$ and 
$\nu^*_1,\nu^*_2\leq2$. 
Because $\chi_{\nu^*_j}$ contains at most one power of $\bar{z}_j$, $ \Psi^{(\nu^*_1,\nu^*_2)}_{3/7}$ has at most $\bar{z}_j^2$ and is therefore restricted to the lowest three LLs.
(We also must have $\nu^*_1,\nu^*_2\geq6/5$ because $(\nu^*_1)^{-1}=4/3-(\nu^*_2)^{-1}\leq5/6$.) We have used the fact that the inverse of the filling factor of a product state is sum of the inverse filling factors of the different factors~\cite{Jain89b}. 
We use here and below the symbol $\chi_{\nu^*}$ to denote a Slater determinant in the standard angular momentum basis in which there is at least one particle in both LLs.
The zero mode (ZM) subspace of $V_\text{TK}$ is spanned by all wave functions of the
form $\Psi^{(\nu^*_1,\nu^*_2)}_{3/7}$~\cite{Bandyopadhyay20}.

Certain other models have been advanced. Bandyopadhyay {\it et al.} have constructed a local two-body interaction for all unprojected Jain wave functions at $n/(2n+1)$~\cite{Bandyopadhyay20} as well as for the Jain parton states ~\cite{Bandyopadhyay18} building upon previous work~\cite{Chen17}. It is not evident, however, how this interaction may be expressed in a real space form or in terms of Haldane pseudopotentials. Anand {\it et al.}~\cite{Anand21,AnandTorus23} 
have introduced an interaction, defined in terms of generalized Haldane pseudopotentials, which does not cause inter-LL scattering, and shown that this interaction can be solved exactly, and its spectrum at $\nu$ has an exact correspondence with that of non-interacting fermions at $\nu^*$, given by $\nu=\nu^*/(2p\nu^*+1)$. In particular, it produces incompressibility at the Jain fractions $\nu=n/(2pn+1)$. 
This formulation also provides a solvable model for non-Abelian FQHE~\cite{Kudo23}.
The eigenfunctions of this interaction, however, are not the unprojected Jain CF states and for $n\geq 2$ have large occupancy of higher LLs.
The 3/7 FQHE has been extensively studied from various other perspectives as well~\cite{Andrews18,Andrews20,Andrews21,Kudo20}.

Our strategy in this paper is to ask if we can add a three-body interaction to the TK interaction to single out the 3/7 wave function $\Phi_3\Phi_1^2$ as the unique zero energy solution. The reason why we can expect a three-body interaction to single out $\Phi_3\Phi_1^2$ from the other TK ZMs is because $\chi_{\nu^*}$ with $\nu^*\geq 1$
has a different behavior than $\Phi_1$ when three particles are brought close to one another: while $\Phi_1$ vanishes as $r^3$, $\chi_{\nu^*}$ vanishes as $r^2$. (A proof is given in Appendix~\ref{appx:Psirr}). As a result, $\Phi_3\Phi_1^2$ vanishes as $r^8$ whereas $\chi_{\nu^*_1}\chi_{\nu^*_2}\Phi_1$ vanishes as $r^7$. The three-body interaction that can take advantage of this difference is 
\begin{equation}
 V_3^{(s,t,u)}=\nabla_1^{2s}\nabla_2^{2t}\nabla_3^{2u}
 \left[\delta^{(2)}(\vec{r}_1-\vec{r}_2)\delta^{(2)}(\vec{r}_1-\vec{r}_3)
 \right], 
 \label{eq:V3body}
 \end{equation}
where $(s,t,u)$ are non-negative integers. With $s+t+u=7$, the wave function $\Phi_3\Phi_1^2$ has a zero expectation value
for this interaction whereas the other states of the form $\chi_{\nu^*_1}\chi_{\nu^*_2}\Phi_1$ do not.

There are several subtle problems with the above argument, which we now mention along with possible resolutions.

{\it Problem 1}: The three-body interaction $ V_3^{(s,t,u)}$ has positive as well as negative eigenvalues. This is problematic because, then, a state with zero 
expectation value is not necessarily an eigenstate, and even if it is, lower energy solutions can exist.
Fortunately, within the Hilbert space defined by the ZMs of $V_\text{TK}$, called the TK-ZM space, the eigenvalues of $ V_3^{(s,t,u)}$ are non-negative. (See discussion below. An analogous situation occurs when one considers spinful fermions with TK interaction. In this case, the TK interaction in general has negative energy solutions. However, if one confines the Hilbert space to the ZMs of the delta function interaction, i.e. to states that vanish when two electrons coincide, then the eigenenergies of the TK interaction are non-negative. See Ref.~\cite{Belkhir93a}.) In what follows, we will first send the coefficient of $V_\text{TK}$ to infinity, so that only the TK-ZM states survive, and then diagonalize $V_3^{(s,t,u)}$ with $s+t+u=7$ within that subspace.  
In other words, our proposed Hamiltonian is
\begin{align}
 H=\lim_{\lambda\rightarrow\infty}
 \lambda V_{\rm TK}+V_3^{(s,t,u)}.
 \label{eq:mainHamiltonian}
\end{align}

{\it Problem 2}: Though 
$\chi_{\nu^*_1}\chi_{\nu^*_2}\Phi_1$ vanishes as $r^7$, 
linear combinations of $\chi_{\nu^*_1}\chi_{\nu^*_2}\Phi_1$ may vanish as $r^8$. In fact, it was shown by Bandyopadhyay {\it et al.} in Ref.~\onlinecite{Bandyopadhyay18} that the Jain unprojected 3/7 state $\Phi_3\Phi_1^2$ can be generated as a linear combination of $\chi_{\nu^*_1}\chi_{\nu^*_2}\Phi_1$. 
This does not rule out the possibility, however, that our model will produce a unique ZM state. 
This can be tested by exact diagonalization (ED) on finite systems.

\begin{figure*}
\includegraphics[width=2\columnwidth]{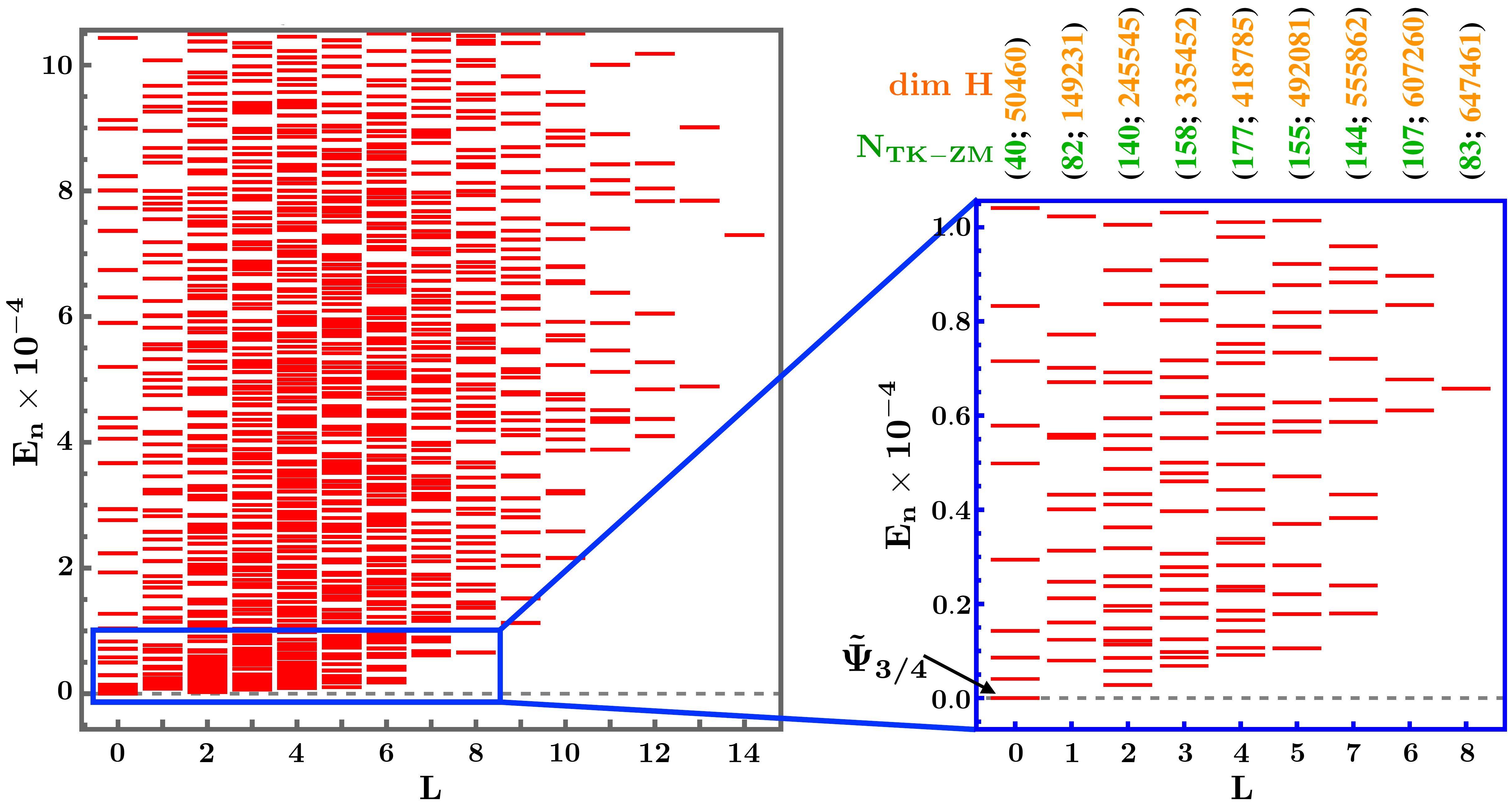}
\caption{\label{fig:nu3_4} Energy spectrum of $V_3^{(0,2,2)}$ 
 within the TK-ZM space for bosons with $(N,2Q)=(9,8)$, which represents  
 filling factor $3/4$. The state $\tilde{\Psi}_{3/4}$ is 
 observed as a unique ZM. The integers colored in orange and green in 
 the right panel indicate the dimension of the Hamiltonian matrix and 
 the TK-ZM space in each $L$ sector, respectively.
 }
\end{figure*}

{\it Problem 3}:  
To perform ED for this Hamiltonian, we will use Haldane's spherical geometry~\cite{Haldane83}, which is the most convenient geometry for dealing with incompressible states. For the 3/7 state, we must have a minimum of 9 particles, because it takes a minimum of 9 particles to fill three LLs to produce $\Phi_3$. The dimension of the Hilbert space for this system is 229,339,157 in the $L_z=0$ sector, which is too large for ED. 
To get around this issue, we study the corresponding bosonic system at $\nu=3/4$. 
For a given particle number $N$, total fluxes $2Q$ for the bosonic system at 
$\nu=3/4$ and the fermionic system at $\nu=3/7$ are related by 
$2Q_\text{boson}=2Q_\text{fermion}-(N-1)$.
The total Hilbert space dimensions across all $L_z$ sectors of the fermionic and bosonic problems are given by $d_{n,F}=^{(2nQ_{\rm fermion}+n^2)}C_{N}$ and  
$d_{n,B}=^{(2nQ_{\rm fermion}+n^2-(n-1)(N-1))}C_{N}$, where $n$ is the number of LLs included. [The ``$r$ choose $s$'' function is defined as $^rC_s=r!/s! (r-s)!$.] From the fact that $^{x}C_y$ is an increasing function of $x$, we see that $d_{n,B}<d_{n,F}$ if more than $1$ LL is included, as in our case (while $d_{n,B}=d_{n,F}$ if only the LLL Hilbert space is allowed). The reduction in the total dimension when going from the fermionic to the bosonic problem is reflected in the individual $L_z$ sectors as well. The dimension of the bosonic system at $N=9$ in the $L_z=0$ sector is given by $12,649,289$ when we include the lowest three LLs.

The arguments about the vanishing properties and interactions remain valid if one replaces the wave functions as well as the two- and three-body interactions as
\begin{align}
 \Psi_{3/7}=\Phi_3\Phi_1^2
 &\rightarrow\tilde{\Psi}_{3/4}=\Phi_3\Phi_1\\
 \Psi_{3/7}^{(\nu_1^*,\nu_2^*)}=\chi_{\nu^*_1}\chi_{\nu^*_2}\Phi_1
 &\rightarrow\tilde{\Psi}_{3/4}^{(\nu_1^*,\nu_2^*)}=\chi_{\nu^*_1}\chi_{\nu^*_2}\\
 V_\text{TK}=\nabla^2_2\delta^{(2)}(\vec{r}_2-\vec{r}_1)
 &\rightarrow \tilde{V}_\text{TK}=\delta^{(2)}(\vec{r}_2-\vec{r}_1)\\
 s+t+u=7 &\rightarrow s+t+u=4,
\end{align}
where the quantities / symbols for bosons 
are marked by a tilde. The specific values of $s,t,u$ will be discussed in later sections.

One may ask how the TK-ZMs of fermions and bosons are related. If a bosonic state $f$ is a ZM of $\tilde{V}_{\rm TK}$, then it is clear that $f\Phi_1$ is a ZM of $V_{\rm TK}$, and linearly independent set of $f$ will produce a linearly independent set of $f\Phi_1$. 
Conversely,
any fermionic ZM of $V_{\rm TK}$, which must be a linear superposition 
of $\chi_{\nu^*_1}\chi_{\nu^*_2}\Phi_1$~\cite{Bandyopadhyay20},
contains the factor $\Phi_1$, which implies that the set of linearly independent ZMs of fermions will produce a set of linearly independent set of ZMs of bosons. This implies that there is an exact one-to-one correspondence between the TK-ZMs of fermions and bosons. Our numerical diagonalization studies presented below are consistent with this statement.

Numerically, for all parameters for which we can study both the fermionic and bosonic systems, we find a one-to-one correspondence 
also between the ZMs of the full Hamiltonian of Eq.~\eqref{eq:mainHamiltonian} that includes both two and three-body interactions.
Assuming this to be generally true, we can address the question of the uniqueness of the fermionic ZM with $N=9$ particles through a study of the corresponding bosonic system, which is computationally more tractable.

{\it Result:} To address these questions, we have considered fermions (bosons) for a range of filling factors between $3/7 \geq \nu > 2/5$ ($3/4 \geq \nu > 2/3$) and numerically diagonalized the three-body interaction within the TK-ZM 
space. We summarize the results here, with details given in subsequent sections.

\begin{itemize}
\item All eigenenergies of Eq.~\eqref{eq:mainHamiltonian} are non-negative.

\item Fermions and bosons at corresponding  filling factors produce the same number of ZMs. 

\item For fermions at $\nu<3/7$, the number of ZMs produced by ED is larger than the number of states of the form $\chi'_{\nu^*}\Phi_1^2$, where $\chi'_{\nu^*}$ is confined within the lowest three LLs. This is an explicit demonstration that many states of the form  $\chi_{\nu_1^*}\chi_{\nu_2^*}\Phi_1$ for fermions are combining to produce ZMs for the 3-body interaction. The same is true for bosons with $\nu<3/4$.

 \item We are able to diagonalize the three-body interaction within the TK-ZM space for bosons at $\nu=3/4$ for 9 particles. We find a unique zero energy state here, which must be the ZM $\Phi_3\Phi_1$. The calculated spectrum is shown in Fig.~\ref{fig:nu3_4}. This implies a unique zero energy state also for 9 fermions at $\nu=3/7$. 

\item The above result is rather nontrivial: there are 40, 82, 140, 158, 177, $\cdots$ ZMs at $L=0$, 1, 2, 3, 4, $\cdots$ for the TK interaction, but when the three-body interaction is turned on, a single ZM remains at $L=0$. This
strongly suggests that our model will produce a unique ZM at $\nu=3/7$ for fermions or at $\nu=3/4$ for bosons for arbitrary number of particles. However, we are not able to prove this statement analytically, nor are we able to perform ED for the next incompressible system which has $N=12$ particles.

\end{itemize}

The plan of the rest of the paper is as follows. In Sec.~\ref{sec:stu},
we discuss how to construct our model interaction $V_3^{(s,t,u)}$ in
greater details. In Sec.~\ref{sec:method}, we give an
explicit expression of our model in the sphere geometry for ED. 
In Sec.~\ref{sec:result}, we discuss other numerical results. Concluding 
remarks are given in Sec.~\ref{sec:conc}. More technical details can be found 
in Appendices.

\section{Model interaction $V_3^{(s,t,u)}$}
\label{sec:stu}

We begin by noting that $ V_3^{(s,t,u)}$ defined in Eq.~\eqref{eq:V3body} satisfies the relation:
\begin{align}
 V_3^{(s,t,u)}= V_3^{(t,s,u)}=\cdots=V_3^{(u,t,s)}.
 \label{eq:stu_perm}
\end{align}
This can be seen by writing 
$\delta^{(2)}(\vec{r}_3-\vec{r}_1)\delta^{(2)}(\vec{r}_3-\vec{r}_2)$ as 
\begin{align}
 \delta_{31}\delta_{32}
 &=\left(
 \delta_{31}\delta_{32}+\delta_{23}\delta_{21}+\delta_{12}\delta_{13}
 \right)/3,
 \label{eq:123}
\end{align}
where $\delta_{ij}$ is a shorthand for $\delta^{(2)}(\vec{r}_i-\vec{r}_j)$.
The expression in Eq.~\eqref{eq:123} is invariant under a permutation of 
particle labels, which leads to Eq.~\eqref{eq:stu_perm}. Hereafter, we set 
$s\leq t\leq u$ without loss of generality.

\subsection{Short distance behavior}
\label{sec:short}
To facilitate the analysis, we use the center of mass coordinate $\vec{R}$ and relative coordinates $\vec{r}_a,\vec{r}_b$ for three particles:
\begin{align}
 \left(
 \begin{array}{l}
  \vec{R} \\
  \vec{r}_a \\
  \vec{r}_b
 \end{array}
 \right)
 =T
 \left(
 \begin{array}{c}
  \vec{r}_1 \\
  \vec{r}_2 \\
  \vec{r}_3
 \end{array}
 \right),\ 
 T=\left(
 \begin{array}{ccc}
  1/3&1/3&1/3 \\
  1&-1&0 \\
  1&0&-1 \\
 \end{array}
 \right).
 \label{eq:Rrarb}
\end{align}
We have $\det T=1$.  
When three particles approach one another, a general wave function of
fermions must vanish at least as $r_ar_b$ 
due to antisymmetrization. 
However, in the LLL, the wave function vanishes faster. Any LLL wave function has the form  $f(\{z_i\})\Phi_1$, and therefore a three-particle wave function 
 vanishes at least as fast as 
\begin{align}
 \Phi_1=\prod_{i<j}(z_i-z_j)\sim r_ar_b^2,\ r_a^2r_b,
 \label{eq:Phi1rrr}
\end{align}
where the notation indicates that the quantity is a linear combination 
of two terms that vanish as $r_ar_b^2$ and $r_a^2r_b$.
When higher LLs are allowed, the availability of nonholomorphic coordinates allows one to construct wave functions vanishing slower, as $r_ar_b$.
As shown in Appendix~\ref{appx:Psirr}, any three-body Slater 
determinant state $\Psi$ where two particles occupy the LLL and the third 
the second LL vanishes as
\begin{align}
 \Psi\sim r_ar_b.
\label{eq:Psirr}
\end{align}
Any Slater determinant  that has a non-zero occupation of such three particle configurations vanishes as $\sim r_ar_b$.
Since the Slater determinant states $\Phi_3$ and 
$\chi_{\nu^*}$ with $\nu^*>1$ contain such three particle configurations, they also scale as 
\begin{align}
 \begin{array}{l}
  \Phi_3\sim r_ar_b,\\
  \chi_{\nu^*}\sim r_ar_b.
 \end{array}
 \label{eq:chirr}
\end{align}
Using Eqs.~\eqref{eq:Phi1rrr} and \eqref{eq:chirr}, we have
\begin{align}
 \begin{split}
  &|\Psi_{3/7}|^2\sim
  r_a^6r_b^{10},\ r_a^7r_b^9,\ldots,\ r_a^{10}r_b^6,
  \\
  &|\Psi_{3/7}^{(\nu_1^*,\nu_2^*)}|^2\sim
  r_a^6r_b^8,\ r_a^7r_b^7,\ r_a^8r_b^6
 \end{split}
 \label{eq:3/7sim}
\end{align}
Analogous behavior follows for bosons:
\begin{align}
 \begin{split}
  &|\tilde{\Psi}_{3/4}|^2\sim
  r_a^4r_b^6,\ r_a^5r_b^5,\ r_a^6r_b^4,\\
  &|\tilde{\Psi}_{3/4}^{(\nu_1^*,\nu_2^*)}|^2
  \sim r_a^4r_b^4.
 \label{eq:3/4sim}
 \end{split}
\end{align}

\subsection{Expectation value of $V_3^{(s,t,u)}$}
\label{sec:EVofV3}
We consider the expectation value of $V_3^{(s,t,u)}$ for a general 
$N$-body wave function $\Psi$:
\begin{align}
\ave{V_3^{(s,t,u)}}_\Psi
\propto&\sum_{i<j<k}
\int d\vec{r}_1\cdots d\vec{r}_N |\Psi|^2
V_3^{(s,t,u)}(\vec{r}_i,\vec{r}_j,\vec{r}_k)\non
\propto&\int d\vec{r}_1\cdots d\vec{r}_N
|\Psi|^2V_3^{(s,t,u)}(\vec{r}_1,\vec{r}_2,\vec{r}_3)\non
=&\int d\vec{r}_1\cdots d\vec{r}_N
\delta_{12}\delta_{13}
\nabla_1^{2s}\nabla_2^{2t}\nabla_3^{2u}|\Psi|^2.
\label{eq:defS}
\end{align}
The goal in this subsection is to find a set of $(s,t,u)$ such that
$\ave{V_3^{(s,t,u)}}_{\Psi_{3/7}}=0$ and 
$\ave{V_3^{(s,t,u)}}_{\Psi_{3/7}^{(\nu_1^*,\nu_2^*)}}\neq0$.

\begin{table}[t!]
 \caption{
 All terms appearing in expansion of
 $\left(\nabla_{a}+\nabla_{b}\right)^{2s}\nabla^{2t}_{a}\nabla^{2u}_{b}$ are
 shown for each $(s,t,u)$. The operators in bold text, denoted $\mathcal{D}$, 
 satisfy
 $\mathcal{D}\left|\Psi_{3/7}^{(\nu^*_1,\nu^*_2)}\right|^2\sim1$ for $s+t+u=7$
 or
 $\mathcal{D}\left|\tilde{\Psi}_{3/4}^{(\nu^*_1,\nu^*_2)}\right|^2\sim1$ for 
 $s+t+u=4$. ``$\circ$'' indicates
 a candidate of $(s,t,u)$ to construct a parent Hamiltonian.
 }
 \label{tab:stu}
 \centering
 \begin{tabular*}{\columnwidth}{@{\extracolsep{\fill}}cccl}
  \toprule
  $s+t+u$ & $(s,t,u)$ & & All terms \\
  \cmidrule(r){1-4}
  7 & $(0,0,7)$ & & $\nabla_a^0\nabla_b^{14}$ \\
  & $(0,1,6)$ & & $\nabla_a^2\nabla_b^{12}$ \\
  & $(0,2,5)$ & & $\nabla_a^4\nabla_b^{10}$ \\
  & $(0,3,4)$ & $\circ$ & $\bm{\nabla_a^6\nabla_b^{8}}$ \\
  & $(1,1,5)$ & & $\nabla_a^2\nabla_b^{12}$, $(\nabla_a\cdot\nabla_b)\nabla_a^2\nabla_b^{10}$, $\nabla_a^4\nabla_b^{10}$ \\
  & $(1,2,4)$ & $\circ$ & $\nabla_a^4\nabla_b^{10}$, $(\nabla_a\cdot\nabla_b)\nabla_a^4\nabla_b^{8}$, $\bm{\nabla_a^6\nabla_b^{8}}$ \\
  & $(1,3,3)$ & $\circ$ & $\bm{\nabla_a^6\nabla_b^{8}}$, $\bm{(\nabla_a\cdot\nabla_b)\nabla_a^6\nabla_b^{6}}$, $\bm{\nabla_a^8\nabla_b^{6}}$\\
  & $(2,2,3)$ & $\circ$ & $\nabla_a^4\nabla_b^{10}$, $(\nabla_a\cdot\nabla_b)\nabla_a^4\nabla_b^{8}$, $\bm{\nabla_a^6\nabla_b^{8}}$, \\
  & & & \qquad $\bm{(\nabla_a\cdot\nabla_b)\nabla_a^6\nabla_b^{6}}$, $\bm{\nabla_a^8\nabla_b^{6}}$\\
  \cmidrule{1-4}
  \morecmidrules
  \cmidrule{1-4}
  4 & $(0,0,4)$ & & $\nabla_a^0\nabla_b^{8}$ \\
  & $(0,1,3)$ & & $\nabla_a^2\nabla_b^{6}$ \\
  & $(0,2,2)$ & $\circ$ & $\bm{\nabla_a^4\nabla_b^{4}}$ \\
  & $(1,1,2)$ & $\circ$ & $\nabla_a^2\nabla_b^{6}$, $(\nabla_a\cdot\nabla_b)\nabla_a^2\nabla_b^{4}$, $\bm{\nabla_a^4\nabla_b^{4}}$\\
  \midrule
  \bottomrule
 \end{tabular*}
\end{table}
To simplify $\ave{V_3^{(s,t,u)}}_\Psi$, we express the derivatives as
\begin{align}
 \nabla_1^{2s}\nabla_2^{2t}\nabla_3^{2u}
 &=
 \left(\nabla_{a}+\nabla_{b}\right)^{2s}
 \nabla^{2t}_{a}
 \nabla^{2u}_{b},
\end{align}
where we have plugged $\nabla_{\vec{R}}=0$ and 
used $(\nabla_1,\nabla_2,\nabla_3)=(\nabla_a+\nabla_b,-\nabla_a,-\nabla_b)$. 
Integrating over $\vec{r}_a$ and $\vec{r}_b$, we get
\begin{align}
 &\int d\vec{r}_1d\vec{r}_2d\vec{r}_3
 \delta_{12}\delta_{13}
 \nabla_1^{2s}\nabla_2^{2t}\nabla_3^{2u}|\Psi|^2\non
 =&\int d\vec{R}\left(\nabla_{a}+\nabla_{b}\right)^{2s}
 \nabla^{2t}_{a}
 \nabla^{2u}_{b}
 |\Psi|^2
 \Bigr|_{\vec{r}_a,\vec{r}_b\to{0}}.
 \label{eq:SIII}
\end{align}
For this to be non-zero for the state $\Psi_{3/7}^{(\nu^*_1,\nu^*_2)}$, we require $s+t+u=7$ as $|\Psi_{3/7}^{(\nu^*_1,\nu^*_2)}|^2\sim r^{14}$ when $r_a,r_b=r\to 0$. In Table~\ref{tab:stu}, we 
list all terms appearing in the expansion of
$\left(\nabla_{a}+\nabla_{b}\right)^{2s}\nabla^{2t}_{a}\nabla^{2u}_{b}$ for different choices of $s,t,u$. 
Comparing them with Eq.~\eqref{eq:3/7sim}, we see that
$\ave{V_3^{(s,t,u)}}_{\Psi_{3/7}}=0$ 
while $\ave{V_3^{(s,t,u)}}_{\Psi_{3/7}^{(\nu^*_1,\nu^*_2)}}$ can be non-zero
if
\begin{align}
 (s,t,u)=(0,3,4),(1,2,4),(1,3,3),(2,2,3).
 \label{eq:stuF}
\end{align}
This makes $V_3^{(s,t,u)}$ with any of these values of $s,t,u$ a candidate parent Hamiltonian. 
For the remaining choices:
\begin{align}
 (s,t,u)=(0,0,7),(0,1,6),(0,2,5),(1,1,5),
 \label{eq:stuF0}
\end{align} 
both $\ave{V_3^{(s,t,u)}}_{\Psi_{3/7}}
$ and $\ave{V_3^{(s,t,u)}}_{\Psi_{3/7}^{(\nu_1^*,\nu_2^*)}}$ are zero. Since $\Psi_{3/7}^{(\nu_1^*,\nu_2^*)}$ spans the TK-ZM space, $V_3^{(s,t,u)}$ becomes a zero matrix in this space.

Applying the above argument to the bosonic $3/4$ problem, we identify 
$V_3^{(s,t,u)}$ with
\begin{align}
 (s,t,u)=(0,2,2),(1,1,2)
 \label{eq:stuB}
\end{align}
as a candidate parent Hamiltonian for $\tilde{\Psi}_{3/4}$; see
Table~\ref{tab:stu}. 

We note here that $\ave{V_3^{(s,t,u)}}_\Psi=0$ does not necessarily lead to 
$\hat{V}_3^{(s,t,u)}\ket{\Psi}=0$. This is guaranteed if $V_3^{(s,t,u)}$ is positive semi-definite. 

In the following sections, we check, by explicit numerical diagonalization, 
whether
our model interaction singles out $\Psi_{3/7}$ or
$\tilde{\Psi}_{3/4}$ from the other TK ZMs.




\section{Matrix elements}
\label{sec:method}
For diagonalization studies, we consider Haldane's spherical 
geometry~\cite{Haldane83},
where $N$ particles move on the surface under a radial magnetic field. The 
total radial flux is $2Q\phi_0$, where $\phi_0=hc/e$ is the flux quantum and
$2Q$ is an integer. Because of rotational symmetry, single-particle states are
labeled by the orbital angular momentum $l$ and its $z$-component $m$. Their
possible values are $l=|Q|,|Q|+1,\ldots$ and $m=-l,-l+1,\ldots,l$.
The $2l+1$ states with $l=|Q|+n$ corresponds to the $n$th LL. Many-body states
are labeled by the total orbital angular momentum $L$. 

In the second-quantized from, the two-body interactions $V_\text{TK}$,
$\tilde{V}_\text{TK}$, and the three-body interaction $V_3^{(s,t,u)}$ are given 
by
\begin{align}
 \mathcal{V}_\text{TK}
 =&\frac{1}{2}\sum_{1,2,1',2'}
 \hat{f}^\dagger_{1}\hat{f}^\dagger_{2}
 V_{12;1'2'}^\text{TK}
 \hat{f}_{2'}\hat{f}_{1'},\\
 \tilde{\mathcal{V}}_\text{TK}
 =&\frac{1}{2}\sum_{1,2,1',2'}
 \hat{b}^\dagger_{1}\hat{b}^\dagger_{2}
 \tilde{V}_{12;1'2'}^\text{TK}
 \hat{b}_{2'}\hat{b}_{1'},\\
 \mathcal{V}_3^{(s,t,u)}
 =&\frac{1}{6}\sum_{1,2,3,1',2',3'}
 \hat{c}^\dagger_{1}\hat{c}^\dagger_{2}\hat{c}^\dagger_{3}
 V_{123;1'2'3'}^{(s,t,u)}
 \hat{c}_{3'}\hat{c}_{2'}\hat{c}_{1'}.
\end{align}
In these expressions, we use shorthands as $f^\dagger_1=f^\dagger_{l_1m_1}$, 
$f^\dagger_{1'}=f^\dagger_{l'_1m'_1}$ and so on, where $\hat{f}^\dagger_{lm}$ 
and $\hat{b}^\dagger_{lm}$ are the creation operators for a fermion and a 
boson, respectively, and $\hat{c}^\dagger_{lm}=\hat{f}^\dagger_{lm}$ or 
$\hat{b}^\dagger_{lm}$. The summation $\sum_{i}$ indicates 
$\sum_{l_i=|Q|}^\infty\sum_{m_i=-l_i}^{l_i}$. The
symbols $V_{12;1'2'}^\text{TK}$, $\tilde{V}_{12;1'2'}^\text{TK}$, and 
$V_{123;1'2'3'}^{(s,t,u)}$ are shorthands for the matrix elements, e.g.,
$(\bra{l_1,m_1}\otimes\bra{l_2,m_2})\hat{V}_\text{TK}(\ket{l'_1,m'_1}\otimes\ket{l'_2,m'_2})$. 
As derived in Appendix~\ref{appx:elements}, they reduce to
\begin{widetext}
 \begin{align}
  V_{12;1'2'}^{(n)}
  =&(-1)^{2Q-m_{12}-m}\sum_{l}
  \left[-l(l+1)\right]^n
  S\left(
  \begin{array}{ccc}
   -Q&Q&0 \\
   l_1&l'_1&l\\
   -m_1&m'_1&m
  \end{array}
  \right)
  S\left(
  \begin{array}{ccc}
   -Q&Q&0 \\
   l_2&l'_2&l\\
   -m_2&m'_2&-m
  \end{array}
  \right).
  \label{eq:1212}\\
  V^{s,t,u}_{123;1'2'3'}
  =&(-1)^{3Q-m_{123}}
  \sum_{l_a,l_b,l_c}
  S^{(s)}\left(
  \begin{array}{ccc}
  -Q&Q&0 \\
   l_1&l'_1&l_a\\
   -m_1&m'_1&m_a
  \end{array}
  \right)
  S^{(t)}\left(
  \begin{array}{ccc}
   -Q&Q&0 \\
   l_2&l'_2&l_b\\
   -m_2&m'_2&m_b
  \end{array}
  \right)
  S^{(u)}\left(
  \begin{array}{ccc}
   -Q&Q&0 \\
   l_3&l'_3&l_c\\
   -m_3&m'_3&m_c
  \end{array}
  \right)
  S\left(
  \begin{array}{ccc}
   0&0&0 \\
   l_a&l_b&l_c\\
   m_a&m_b&m_c
  \end{array}
  \right),
  \label{eq:123123}
 \end{align}
\end{widetext}
where 
\begin{align}
 &S^{(t)}\left(
  \begin{array}{ccc}
   Q_1&Q_2&Q_3 \\
   l_1&l_2&l_3\\
   m_1&m_2&m_3
  \end{array}
 \right)
 \equiv
 \left[-l_3(l_3+1)\right]^{t}
 S\left(
  \begin{array}{ccc}
   Q_1&Q_2&Q_3 \\
   l_1&l_2&l_3\\
   m_1&m_2&m_3
  \end{array}
 \right),\non
 &S\left(
  \begin{array}{ccc}
   Q_1&Q_2&Q_3 \\
   l_1&l_2&l_3\\
   m_1&m_2&m_3
  \end{array}
 \right)
 \equiv\int d\vec{\Omega}Y_{Q_1l_1m_1}Y_{Q_2l_2m_2}Y_{Q_3l_3m_3},\nonumber
\end{align}
and $Y_{Qlm}$ is the monopole harmonics. 
Here, $V^\text{TK}_{12;1'2'}=V^{(1)}_{12;1'2'}$ and 
$\tilde{V}^\text{TK}_{12;1'2'}=V^{(0)}_{12;1'2'}$.
In Eq.~\eqref{eq:1212}, 
$m_{12}\equiv m_1+m_2$, and 
$m\equiv m_1-m_1'$. In Eq.~\eqref{eq:123123}, 
$m_{123}\equiv m_1+m_2+m_3$,
$m_a=m_1-m_1'$, $m_b=m_2-m_2'$, $m_c=-m_a-m_b$.
The range of each summation is explicitly given in 
Appendix~\ref{appx:elements}.

The interactions $\mathcal{V}_\text{TK}$, $\tilde{\mathcal{V}}_\text{TK}$, and
$\mathcal{V}_3^{(s,t,u)}$ conserve the total orbital angular momentum $L$ and its
$z$-component $L_z$. Within the subspace specified by $L_z$, we diagonalize
these interactions using the Lanczos method. The Hilbert space is restricted in
the lowest three LLs. 
In the following, we focus on $V_3^{(0,3,4)}$ for fermions
and $V_3^{(0,2,2)}$ for bosons as representatives of Eqs.~\eqref{eq:stuF} and 
\eqref{eq:stuB}.

\section{Exact diagonalization results}
\label{sec:result}

\subsection{Pseudopotentials}
\label{sec:PPs}
We first ask if the three-body interaction $V_3^{(s,t,u)}$ is positive 
semi-definite or not. To this end, we calculate pseudopotentials 
$V_M$~\cite{Haldane83} 
by diagonalizing the interactions for three particles, where $M\equiv3Q-L$ 
corresponds to the relative angular momentum in the disk geometry.

\begin{figure}
\includegraphics[width=\columnwidth]{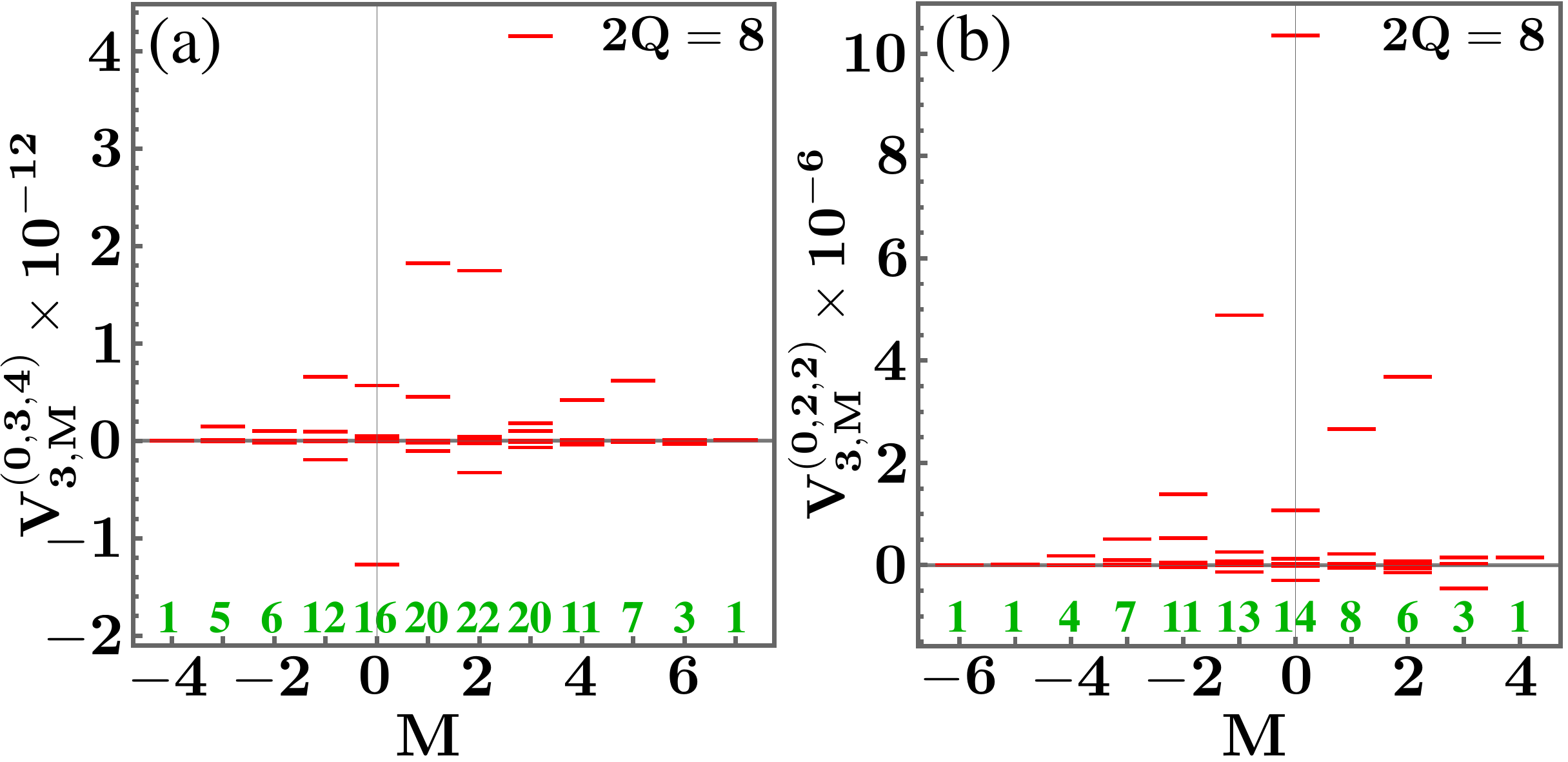}
 \caption{\label{fig:PPs}
 Pseudopotentials of (a) $V_3^{(0,3,4)}$ for fermions and (b) 
$V_3^{(0,2,2)}$ for bosons. 
 Only the nonzero pseudopotentials are shown for simplicity. At each $M$, the number of 
 linearly independent states with non-zero energy is shown in green. 
 Note that some pseudopotentials are 
 nearly degenerate on the scale shown; for example, it may appear that we have
 only two states at $M=-3$ in (a) but in reality there are five. We have set 
 $2Q=8$ in both figures.
 [We have confirmed that both positive and negative values of the pseudopotentials are also obtained for the system in (a) when it is confined to the 
lowest {\it two} LLs.]
 }
\end{figure}
Figure~\ref{fig:PPs} shows the pseudopotentials (energy eigenvalues of a three-particle system)
$V_{3,M}^{(0,3,4)}$ for fermions and $V_{3,M}^{(0,2,2)}$ 
for bosons.
Only pseudopotentials with nonzero values are shown for simplicity.  Both positive and negative values 
are obtained in the two figures.
After projecting into the TK-ZM space, we are left with many fewer 
pseudopotentials, which are all non-negative.
Table~\ref{tab:PPs} summarizes those numbers.
There are only three
nonzero 
pseudopotentials in each case and they are all positive. Furthermore, the 
values increase with $2Q$, which suggests that $V_3^{(0,3,4)}$ 
for fermions and $V_3^{(0,2,2)}$ for bosons are positive 
semi-definite within the TK-ZM space for 
arbitrary $2Q$.
[With $(N,2Q)=(5,8)$ for fermions, we confirm that 
$\lambda V_\text{TK}+V_3^{(0,3,4)}$ with finite $\lambda$ yields no ZMs while 
$V_3^{(0,3,4)}$ within the TK-ZM space does yield ZMs as shown in 
Table~\ref{tab:counting} below.]
This guarantees that $\Psi_{3/7}$ and 
$\tilde{\Psi}_{3/4}$, which give the zero expectation values for each
of the interactions considered, are zero-energy eigenstates.
We investigate the question of their uniqueness below.
\begin{table}[t!]
 \caption{
 Nonzero pseudopotentials of $V_3^{(0,3,4)}$ for fermions and 
$V_3^{(0,2,2)}$ for bosons within the TK-ZM space.
 }
 \label{tab:PPs}
 \centering
 \begin{tabular*}{\columnwidth}{@{\extracolsep{\fill}}cccc}
  \toprule
  & \multicolumn{3}{c}{$V^{(0,3,4)}_{3,M}$} \\
  \cmidrule(r){2-4}
  $2Q$ & $M=1$ & $2$ & $3$ \\
  \cmidrule{1-4}
  \morecmidrules
  \cmidrule{1-4}
  6 & $4.30344\times10^{5}$ & $1.25378\times10^{7}$ & $2.17974\times10^{8}$ \\
  7 & $1.00218\times10^{6}$ & $3.09694\times10^{7}$ & $5.79790\times10^{8}$ \\
  8 & $2.18070\times10^{6}$ & $7.06699\times10^{7}$ & $1.40204\times10^{9}$ \\
  \cmidrule{1-4}
  \morecmidrules
  \cmidrule{1-4}
  & \multicolumn{3}{c}{$V^{(0,2,2)}_{3,M}$} \\
  \cmidrule(r){2-4}
  $2Q$ & $M=-2$ & $-1$ & $0$ \\
  \cmidrule{1-4}
  \morecmidrules
  \cmidrule{1-4}
  6 & $1.54430\times10^2$ & $2.88793\times10^{3}$ & $3.27657\times10^{4}$ \\
  7 & $2.65699\times10^2$ & $5.10736\times10^{3}$ & $5.99266\times10^{4}$ \\
  8 & $4.37421\times10^2$ & $8.60243\times10^{3}$ & $1.03738\times10^{5}$ \\
  \midrule
  \bottomrule
 \end{tabular*}
\end{table}

\subsection{Short distance behavior of $\chi_{\nu^*}$}
\label{sec:chi}
We noted previously that any Slater determinant $\chi_{\nu^*}$ 
with $\nu^*>1$
vanishes as $\sim r_ar_b$ [see Eq.~\eqref{eq:chirr}] when three particles come close to one another. It is possible to construct general wave functions for fermions at $\nu>1$ which vanish as $\sim r_ar_b^2,r_a^2r_b$.
An explicit example is 
\begin{align}
 \psi
 &=\mathcal{A}[(\bar{z}_1-\bar{z}_2)(z_1-z_3)(z_1-z_4)(z_1-z_5)\cdots(z_1-z_N)\non
 &\qquad\qquad\qquad\,
 (z_2-z_3)(z_2-z_4)(z_2-z_5)\cdots(z_1-z_N)\non
 &\qquad\qquad\qquad\qquad\qquad
 (\bar{z}_3-\bar{z}_4)(z_3-z_5)\cdots(z_1-z_N)\non
 &\qquad\qquad\qquad\qquad\qquad\qquad\qquad\qquad\quad\ \ 
 \cdots\qquad\qquad]\non
 &=\mathcal{A}\left[\prod_{i<j}(z_i-z_j)
 \prod_{i\in\text{odd}}
 \frac{(\bar{z}_{i}-\bar{z}_{i+1})}{(z_{i}-z_{i+1})}\right].
\end{align}
The filling factor of this wave function is close to unity. Of course, for $\nu<1$ we can construct wave functions of the type $\prod_{i<j}(z_i-z_j)f$ where $f$ is a symmetric polynomial of $z_j$ and $\bar{z}_j$ but with no more than one power of $\bar{z}_j$. 
In this subsection, we numerically show that it is not possible to construct such wave functions for $\nu^*>6/5$, i.e. an arbitrary wave function (i.e. any linear superposition of Slater determinants) at $\nu^*\geq6/5$
vanishes as $\sim r_ar_b$.

To see this, we use the three-body interaction $V_3^{(0,1,1)}$. 
As shown in Table~\ref{tab:011}, $V_3^{(0,1,1)}$ for fermions is 
positive semidefinite and, thus, states that vanish as $\sim r_ar_b^2,r_a^2r_b$
are obtained as its ZMs. In Table~\ref{tab:chi}, we list 
the number of ZMs of $V_3^{(0,1,1)}$ in fermionic systems with 
various $(N,2Q)$.
We find empirically that there is no ZM if
\begin{align}
 2Q^*<\left\{
 \begin{array}{ll}
  N-2& \text{for odd $N$,} \\
  N-3& \text{for even $N$.} \\
  \end{array}
 \right.
 \label{eq:2Qstar}
\end{align}
\begin{table}[t!]
 \caption{
 Nonzero pseudopotentials of $V_3^{(0,1,1)}$ for fermions.
 }
 \label{tab:011}
 \centering
 \begin{tabular*}{\columnwidth}{@{\extracolsep{\fill}}ccc}
  \toprule
  & \multicolumn{2}{c}{$V_{3,M}^{(0,1,1)}$} \\
  \cmidrule(r){2-3}
  $2Q$ & $M={-1}$ & $M=0$ \\
  \cmidrule{1-3}
  \morecmidrules
  \cmidrule{1-3}
  1 & 0.182378 & 1.77819 \\
  2 & 0.422172 & 4.66801 \\
  3 & 0.863533 & 10.3624 \\
  4 & 1.59581 & 20.2545 \\
  \midrule
  \bottomrule
 \end{tabular*}
\end{table}
\begin{table}[t]
 \caption{
 Number of ZMs of $V_3^{(0,1,1)}$ for fermions.
 ``ZM'' indicates the existence of one or more ZMs, although
 their number is unknown. 
 }
 \label{tab:chi}
 \centering
 \begin{tabular*}{\columnwidth}{@{\extracolsep{\fill}}cccccccccccc}
  \toprule
  & \multicolumn{11}{c}{Number of ZM of $V_3^{(0,1,1)}$} \\
  \cmidrule(r){2-12}
  $2Q$ & $N=3$ & 4 & 5 & 6 & 7 & 8 & 9 & 10 & 11 & 12 & 13\\
  \cmidrule{1-12}
  \morecmidrules
  \cmidrule{1-11}
  1 & 3 & 1 & 0 \\
  2 & 10 & 6 & 0 & 0 & 0 \\
  3 & 17 & 22 & 7 & 1 & 0 & 0 \\
  4 & 28 & 47 & 40 & 10 & 0 & 0 & 0 \\
  5 & & & 108 & 74 & 13 & 1 & 0 & 0 \\
  6 & & & & 242 & 124 & 19 & 0 & 0 & 0 \\
  7 & & & & & 505 & 208 & 22 & 1 & 0 & 0 \\
  8 & & & & & & ZM & ZM & ZM & 0 & 0 & 0 \\
  \midrule
  \bottomrule
 \end{tabular*}
\end{table}

Recall that in $\Psi_{3/7}^{(\nu^*_1,\nu^*_2)}=\chi_{\nu_1^*}\chi_{\nu_2^*}\Phi_1$ 
we must have $\nu_1^*,\nu_2^*\geq6/5$. 
In the spherical geometry, the constraints $1^{-1}+(\nu^*_1)^{-1}+(\nu^*_2)^{-1}=(3/7)^{-1}$ 
and $\nu^*_1,\nu^*_2\leq2$ translate into 
\begin{align}
 &(N-1)+2Q_1^*+2Q_2^*=\frac{7N}{3}-5\\
 &(2Q^*_j+1)+(2Q^*_j+3)\geq N,\text{ $i=1,2$,}
\end{align}
where $2Q_j^*$ is the flux corresponding to $\nu_j^*$. These lead to 
\begin{align}
 2Q_j^*\leq5N/6-2
\end{align} 
since $2Q_1=4N/3-4-2Q_2^*\leq4N/3-4-(N/2-2)=5N/6-2$.
This is satisfied by Eq.~\eqref{eq:2Qstar} 
for any $N\geq9$. Because the 3/7 state has $N\geq 9$, this implies that there is no ZM of the product form
$\phi_{\nu_1^*}\phi_{\nu_2^*}\Phi_1$, where $\phi_{\nu_j^*}$ is an arbitrary state (as opposed to a single 
Slater determinant state) at $\nu_j^*\geq 6/5$.
Analogous result holds for bosons. 

This, however, does not rule out that linear superpositions
of product states of the type 
$\chi_{\nu_1^*}\chi_{\nu_2^*}\Phi_1$ may vanish as $\sim r_ar_b^2, r_a^2r_b$. In fact, we 
already know that this is possible, as $\Phi_3 \Phi_1^2$ can be expressed as a linear superposition 
of such states~\cite{Bandyopadhyay18}. The key question is whether that is the only such state or there is more than 
one such state. We address this by direct numerical diagonalization in the next section.

\subsection{Numerical diagonalization}
\label{sec:unique}

We perform ED for the model interaction $V_3^{(0,2,2)}$ for $5\leq N\leq 9$ bosons in the range $3/4\geq \nu>2/3$, and for the model interaction $V_3^{(0,3,4)}$ for $5\leq N\leq 7$ fermions in the range $3/7\geq \nu >2/5$. 
The number of ZMs for these systems is shown in Table~\ref{tab:counting} as a function of the total orbital angular momentum $L$. We find that the number of ZMs is identical for the corresponding bosonic and fermionic systems for $5\leq N\leq 7$ where both systems are diagonalizable, strongly suggesting that the number of ZMs of corresponding bosonic and fermionic systems are equal in general. Though we discuss only the spectrum for $(s,t,u)=(0,3,4)$ for fermions, calculations in specific finite systems suggest that the remaining candidates for $(s,t,u)$ given in Eq.~\eqref{eq:stuF} produce the same ZM counting.

The bosonic system at $\nu=3/4$ 
requires a minimum of $N=9$ particles, which is the largest system size that we can currently diagonalize.
As shown in Table~\ref{tab:counting}, as well as in the full energy spectrum in 
Fig.~\ref{fig:nu3_4}, a {\it unique} ZM with $L=0$ is obtained here, which must be  
$\tilde{\Psi}_{3/4}$. The discussion in the preceding paragraph 
 implies a unique ZM for the interaction $V_3^{(0,3,4)}$ for $N=9$ fermions at
$\nu=3/7$. As remarked in the Introduction section, the non-triviality of the result suggests that 
our model interaction very likely produces a unique ZM for bosons (fermions) at $\nu=3/4$ ($\nu=3/7$) 
 for arbitrary $N$. (The next bosonic system at $\nu=3/4$ has $N=12$ particles with $2Q=12$; this system has 12,982,724,934 basis states with $L_z=0$ for which exact diagonalization is currently not feasible.)
\begin{table*}[]
 \caption{Number of ZMs of $V_3^{(0,3,4)}$ for fermions 
 and $V_3^{(0,2,2)}$ for bosons within the TK-ZM space as a function of the total orbital angular momentum $L$. For $N=5,6,7$ we obtain the number of ZMs for both bosons and fermions and find identical numbers; for $N=8,9$ our calculations are for bosons only.
 The quantity $\dim H$ is the  dimension of the full Hilbert space with $L_z=0$.
 $N_\text{TK-ZM}$ is the number of TK-ZMs with $L_z=0$.
 We also evaluate the number of ZMs of the type given in Eq.~\eqref{eq:PsiAB}, which is shown in parentheses whenever it is different from the actual number of ZMs.
 }
 \label{tab:counting}
 \centering
 \begin{tabular*}{2.07\columnwidth}{@{\extracolsep{\fill}}ccccccccccccccccccccc}
  \toprule
  & \multicolumn{2}{c}{Fermions} & \multicolumn{2}{c}{Bosons} & \multicolumn{15}{c}{$L$}\\
  \cmidrule(r){2-3}\cmidrule(r){4-5}\cmidrule(r){7-21}
  $N$ & $2Q$ & $\dim H$ & $2Q$ & $\dim H$ & $N_\text{TK-ZM}$ & $0$ & 1 & 2 & 3 & 4 & 5 & 6 & 7 & 8 & 9 & 10 & 11 & 12 & 13 & 14 \\
  \cmidrule{1-21}
  \morecmidrules
  \cmidrule{1-21}
  5 & 8 & 13,442 & 4 & 3,956 & 138 & 8(7) & {11} & 16(15) & {10} & {5} & {0} & {0} & {0} & {0} & - & - & - & - & - & - \\
  6 & 10 & 145,079 & 5 & 28,480 & 258 & 3(2) & 14(13) & 8(7) & {9} & {1} & {0} & {0} & {0} & {0} & {0} & - & - & - & - & - \\
  7 & 12 & 1,637,730 & 6 & 212,166 & 454 & 1(0) & 7(4) & 3(2) & {2} & {0} & {0} & {0} & {0} & {0} & {0} & {0} & {0} & - & - & - \\
  8 & - & - & 7 & 1,621,444 & 761 & 3(2) & {1} & {1} & {0} & {0} & {0} & {0} & {0} & {0} & {0} & {0} & {0} & {0} & - & - \\
  9 & - & - & 8 & 12,649,289 & 1203 & {1} & {0} & {0} & {0} & {0} & {0} & {0} & {0} & {0} & {0} & {0} & {0} & {0} & {0} & {0} \\
  \midrule
  \bottomrule
 \end{tabular*}
\end{table*}

We end by presenting an observation on the form of 
the ZM states of our model interaction,  
focusing on the fermionic ZMs at $3/7\geq\nu>2/5$; translation 
to bosons is straightforward.
Two types of product ZM states can be readily constructed:
\begin{align}
 \begin{split}
  &\Psi_{\nu}^\text{A}
  =\chi_{\nu^*}'\Phi_1^2,\\
  &\Psi_\nu^\text{B}
  =\chi_{\nu^*_1}\phi_{\nu^*_2}\Phi_1,
 \end{split}
 \label{eq:PsiAB}
\end{align}
where $\chi_{\nu^*}'$ is a Slater determinant 
{confined to the lowest three LLs}
and $\phi_{\nu^*}$ is a linear combination of Slater determinants 
within the lowest two LLs that vanishes as $\sim r_ar_b^2,r_a^2r_b$. 
The state $\phi_{\nu^*}$ can be constructed by diagonalizing $V_3^{(0,1,1)}$.
Both
$\Psi_{\nu}^\text{A}$ and $\Psi_{\nu}^\text{B}$ are ZMs of our model 
interaction
as the three factors within them scale as $r^2$, $r^3$ and $r^3$ when three particles approach each other. 
We ask if all ZMs of our model interaction belong to these two types. For $\nu<3/7$, the number of linearly independent states of the forms $\Psi_{\nu}^\text{A}$ and $\Psi_{\nu}^\text{B}$ (shown in parenthesis in the Table~\ref{tab:counting}) is always less than the number of ZMs produced by ED, demonstrating that there also exist ZMs that are not of the above product form but linear combinations of $\chi_{\nu^*_1}\chi_{\nu^*_2}\Phi_1$.
The number of such additional ZMs decreases as the $3/4$ bosonic state with $N=9$ is approached, eventually vanishing at $3/4$ bosonic state with $N=9$.



\section{Concluding Remarks}
\label{sec:conc}
In summary, we have constructed a candidate parent Hamiltonian for the 
unprojected Jain wave function at $\nu=3/7$ for fermions or
at $\nu=3/4$ for bosons. This model consists of an infinitely strong two-body Trugman-Kivelson 
interaction, plus a three-body interaction. We have numerically demonstrated that our model
produces a unique zero energy ground state for 9 particles. We believe
this to be the case for arbitrary particle numbers although we have not 
succeeded in proving that analytically.

As noted above, the unprojected Jain $2/5$ state $\Phi_2\Phi_1^2$ is the unique ground state of the TK interaction within the Hilbert space of the lowest two LLs. One may ask if it remains the unique zero energy ground state when we add a three-body interaction, such as $V_3^{(0,3,4)}$. Indeed, this three-body term has zero expectation value with respect to the wave function $\Phi_2\Phi_1^2$. However,  many pseudopotentials have negative energies for this three-body interaction, as explicitly noted in Fig.~\ref{fig:PPs}. As a result, $\Phi_2\Phi_1^2$ remains a unique zero energy {\it ground state} for $\lambda V_{TK}+ V_3^{(0,3,4)}$ only in the limit $\lambda\rightarrow \infty$. For general values of $\lambda$, $\Phi_2\Phi_1^2$ has zero energy, but it is not the ground state.

It is natural to ask if our strategy can be applied to other unprojected Jain wave functions at 
$\nu=n/(2pn+1)$. 
The wave function $\Phi_n\Phi_1^{2p}$ 
vanishes as $r^{2p+1}$ when two particles are brought close to one another
and thus has zero expectation value for a generalized TK interaction
$V_\text{TK}^{(s)}=\nabla^{2s}_2\delta^{(2)}(\vec{r}_2-\vec{r}_1)$ with
$s=2p-1$. Furthermore, as proved in Appendix~\ref{appx:Psirr}, the
Slater determinant state $\Phi_{n\geq2}$ vanishes as $r^2$ when three particles
approach one another, while $\Phi_1$ vanishes as $r^3$. Therefore, $\Phi_n\Phi_1^{2p}$
vanishes as $r^{6p+2}$ and has zero expectation value for $V_3^{(s,t,u)}$ with
$s+t+u=6p+1$. If necessary, one can generalize these interactions to an 
$N$-body one as
$V^{(s_1,\ldots,s_N)}_N
=\nabla_1^{2s_1}\cdots\nabla_N^{2s_N}\left[
\delta^{(2)}(\vec{r}_1-\vec{r}_2)\cdots(\vec{r}_1-\vec{r}_N)
\right]$ 
by investigating behaviors of $N$ particles in a target state. 
Whether these models produce the unprojected Jain wave functions as the {\it unique} zero energy ground states will 
require a more detailed investigation, along the lines presented above for $\nu=3/7$. 

Finally, we note that this method is not useful for the negative-flux Jain states at $\nu=n/(2pn-1)$, because the unprojected wave functions $\Phi_n^*\Phi_1^{2p}$ have arbitrary large powers of $\bar{z}_j$ and hence a nonzero occupation of an infinite number of LLs in the thermodynamic limit. 

\begin{acknowledgments}
 K.K. thanks JSPS for support from Overseas Research Fellowship.
 A.S. and J.K.J acknowledge financial support from the U.S. National Science 
 Foundation under grant no. DMR-2037990. We acknowledge Advanced CyberInfrastructure computational resources 
 provided by The Institute for CyberScience at The Pennsylvania State 
 University. 
 We thank National Supercomputing Mission (NSM) for providing computing 
 resources of 'PARAM Brahma' at IISER Pune, which is implemented by C-DAC and 
 supported by the Ministry of Electronics and Information Technology (MeitY) 
 and Department of Science and Technology (DST), Government of India. SGJ thanks TIFR, Mumbai for their hospitality during the completion of this work. 
\end{acknowledgments}

\appendix
\section{Proof of Eq.~\eqref{eq:Psirr}}
\label{appx:Psirr}
Single-particle states in the LLL and the second LL are given by
\begin{align}
  \begin{array}{ll}
   \eta_{0,m}\propto z^m & (m=0,1,\ldots),
    \label{eq:0m}\\
   \eta_{1,m}\propto z^m(2m+2-z\bar{z}) & (m=-1,0,\ldots),
  \end{array}
\end{align}
where $z=x-iy$. The angular momentum of each state is $m$. By using coordinates
defined in Eq.~\eqref{eq:Rrarb}, that is,
\begin{align}
 &z_1=Z+\frac{z_a}{3}+\frac{z_b}{3},\\
 &z_2=Z-\frac{2z_a}{3}+\frac{z_b}{3},\\
 &z_3=Z+\frac{z_a}{3}-\frac{2z_b}{3},
\end{align}
a three-body Slater 
determinant $\Psi=\det{\eta_{0,m},\eta_{0,n},\eta_{1,k}}$ is evaluated as 
\begin{widetext}
 \begin{align}
 \Psi
 &\propto
  \left|
  \begin{array}{ccc}
   z_1^m&z_2^m&z_3^m \\
   z_1^n&z_2^n&z_3^n \\
   z_1^{k+1}\bar{z}_1&z_2^{k+1}\bar{z}_2&z_3^{k+1}\bar{z}_3
  \end{array}
  \right|+O_3(z_a,z_b) \non
  &=
  \left|
  \begin{array}{ccc}
   \frac{z_1^m+z_2^m+z_3^m}{3} & z_1^m - z_2^m & z_1^m - z_3^m \\
   \frac{z_1^n+z_2^n+z_3^n}{3} & z_1^n - z_2^n & z_1^n - z_3^n \\
   \frac{z_1^{k+1}\bar{z}_1+z_2^{k+1}\bar{z}_2+z_3^{k+1}\bar{z}_3}{3} & 
    z_1^{k+1}\bar{z}_1-z_2^{k+1}\bar{z}_2 &
    z_1^{k+1}\bar{z}_1-z_3^{k+1}\bar{z}_3 \\
  \end{array}
  \right|+O_3(z_a,z_b) \non
  &=
  \left|
  \begin{array}{ccc}
   Z^m & mZ^{m-1}z_a & mZ^{m-1}z_b \\
   Z^n & nZ^{n-1}z_a & nZ^{n-1}z_b \\
   Z^{k+1} & Z^k\left(Z\bar{z}_a+(k+1)\bar{Z}z_a\right) 
    & Z^k\left(Z\bar{z}_a+(k+1)\bar{Z}z_a\right) \\
  \end{array}
  \right|+O_3(z_a,z_b) \non
  &=
  Z^{m-1+n-1+k}\left|
		  \begin{array}{ccc}
		   \left(1-\frac{m}{n}\right)Z & 0 & 0 \\
		   Z & nz_a & nz_b \\
		   1 & Z\bar{z}_a+(k+1)\bar{Z}z_a
		    & Z\bar{z}_a+(k+1)\bar{Z}z_a \\
		  \end{array}
  \right|+O_3(z_a,z_b) \non
  &=
  Z^{m+n+k}(n-m)\left|
  \begin{array}{cc}
   z_a & z_b \\
   \bar{z}_a & \bar{z}_a \\
  \end{array}
  \right|+O_3(z_a,z_b),
 \end{align}
\end{widetext}
where $O_n(z_a,z_b)$ represents a polynomial of $z_a$ and $z_b$ where the 
sum of their powers in each term is greater than $n$.

\section{Derivation of Eqs.~\eqref{eq:1212} and \eqref{eq:123123}}
\label{appx:elements}

We use the following properties of the monopole harmonics 
$Y_{Q,l,m}$~\cite{Wu77,Jain07}:
\begin{align}
 \delta^{(2)}(\vec{r}_2-\vec{r}_1)
 &=\sum_{l=0}^\infty\sum_{m=-l}^l
 Y_{l,m}^*(\vec{\Omega}_2)Y_{l,m}(\vec{\Omega}_1),\\
 \nabla^2Y_{l,m}
 &=-l(l+1)Y_{l,m},\\
 Y_{l,m}^*
 &=(-1)^{-m}Y_{l,-m},\\
 Y_{l_am_a}Y_{l_bm_b}&=(-1)^{m_c}\times \non
 &\sum_{l_c=|l_a-l_b|}^{l_a+l_b}
 S\left(
 \begin{array}{ccc}
  0&0&0 \\
  l_a&l_b&l_c\\
  m_a&m_b&m_c
 \end{array}
 \right)
 Y_{l_c,-m_c},
\end{align}
where $Y_{l,m}\equiv Y_{Q,l,m}$, $m_c=-m_a-m_b$.

\subsection{Two-body interaction}
To treat $V_\text{TK}$ and $\tilde{V}_\text{TK}$ simultaneously, we consider the
following two-body interaction:
\begin{align}
 V_2^{(n)}
 =&\nabla_2^{2n}\delta^{(2)}(\vec{r}_2-\vec{r}_1)\non
 =&\nabla_2^{2n}\sum_{l=0}^\infty
 \sum_{m=-l}^l(-1)^{-m}Y_{l,-m}(\vec{\Omega}_2)Y_{l,m}(\vec{\Omega}_1)\non
 =&\sum_{l=0}^\infty\left[-l(l+1)\right]^n
 \sum_{m=-l}^l(-1)^{-m}Y_{l,-m}(\vec{\Omega}_2)Y_{l,m}(\vec{\Omega}_1)
\end{align}
Note that $V_\text{TK}=V_2^{(1)}$ and $\tilde{V}_\text{TK}=V_2^{(0)}$. 
The matrix element is given by
\begin{widetext}
\begin{align}
 &(\bra{l_1,m_1}\otimes\bra{l_2,m_2})\hat{V}_2^{(n)}
 (\ket{l'_1,m'_1}\otimes\ket{l'_2,m'_2})\non
 =&\sum_{l=0}^\infty
 \left[-l(l+1)\right]^n\sum_{m=-l}^l
 (-1)^{-m}
 \int d\vec{\Omega}_1d\vec{\Omega}_2
 \left[
 Y_{Q,l_2,m_2}^*(\vec{\Omega}_2)
 Y_{Q,l_1,m_1}^*(\vec{\Omega}_1)
 \right]
 \left[
 Y_{l,-m}(\vec{\Omega}_2)Y_{l,m}(\vec{\Omega}_1)
 \right]
 \left[
 Y_{Q,l'_1,m'_1}(\vec{\Omega}_1)
 Y_{Q,l'_2,m'_2}(\vec{\Omega}_2)
 \right]\non
 =&\sum_{l=0}^\infty
 \left[-l(l+1)\right]^n\sum_{m=-l}^l
 (-1)^{2Q-m_2-m_1-m}
 S\left(
 \begin{array}{ccc}
  -Q&Q&0 \\
  l_1&l'_1&l\\
  -m_1&m'_1&m
 \end{array}
 \right)
 S\left(
 \begin{array}{ccc}
  -Q&Q&0 \\
  l_2&l'_2&l\\
  -m_2&m'_2&-m
 \end{array}
 \right)\non
 =&
 (-1)^{2Q-m_2-m_1-m}\sum_{l=\max\left[
 |l_1-l'_1|,|l_2-l'_2|\right]}
 ^{\min\left[
 l_1+l'_1,l_2+l'_2\right]}
 \left[-l(l+1)\right]^n
 S\left(
 \begin{array}{ccc}
  -Q&Q&0 \\
  l_1&l'_1&l\\
  -m_1&m'_1&m
 \end{array}
 \right)
 S\left(
 \begin{array}{ccc}
  -Q&Q&0 \\
  l_2&l'_2&l\\
  -m_2&m'_2&-m
 \end{array}
 \right)
 \Biggr|_{m=m_1-m'_1}.
\end{align}
One gets Eq.~\eqref{eq:1212} by defining $m_1+m_2$ as $m_{12}$.

\subsection{Three-body interaction}
We can write
\begin{align}
 &\delta^{(2)}(\vec{r}_3-\vec{r}_1)\delta^{(2)}(\vec{r}_3-\vec{r}_2)\non
 =&\sum_{l_a,l_b=0}^\infty
 \sum_{m_a=-l_a}^{l_a}\sum_{m_b=-l_b}^{l_b}
 Y^*_{l_am_a}(\vec{\Omega}_3)Y_{l_am_a}(\vec{\Omega}_1)
 Y^*_{l_bm_b}(\vec{\Omega}_3)Y_{l_bm_b}(\vec{\Omega}_2)\non
 =&
 \sum_{l_a,l_b=0}^\infty
 \sum_{m_a=-l_a}^{l_a}\sum_{m_b=-l_b}^{l_b}
 \left[
 (-1)^{m_c}\sum_{l_c=|l_a-l_b|}^{l_a+l_b}
 S\left(
 \begin{array}{ccc}
  0&0&0 \\
  l_a&l_b&l_c\\
  m_a&m_b&m_c
 \end{array}
 \right)
 Y_{l_c,-m_c}(\vec{\Omega}_3)
 \right]^*
 Y_{l_am_a}(\vec{\Omega}_1)Y_{l_bm_b}(\vec{\Omega}_2),\non
 =&
 \sum_{l_a,l_b=0}^\infty
 \sum_{l_c=|l_a-l_b|}^{l_a+l_b}
 \sum_{m_a=-l_a}^{l_a}\sum_{m_b=-l_b}^{l_b}
 S\left(
 \begin{array}{ccc}
  0&0&0 \\
  l_a&l_b&l_c\\
  m_a&m_b&m_c
 \end{array}
 \right)
 Y_{l_a,m_a}(\vec{\Omega}_1)
 Y_{l_bm_b}(\vec{\Omega}_2)Y_{l_cm_c}(\vec{\Omega}_3),
\end{align}
where $m_c=-m_a-m_b$. Using this, we have
\begin{align}
 \hat{V}_3^{(s,t,u)}
 &=\sum_{l_a,l_b=0}^\infty
 \sum_{l_c=|l_a-l_b|}^{l_a+l_b}
 \left[-l_a(l_a+1)\right]^{s}\left[-l_b(l_b+1)\right]^{t}
 \left[-l_c(l_c+1)\right]^{u}\times\non
 &\qquad\qquad\qquad\qquad
 \sum_{m_a=-l_a}^{l_a}\sum_{m_b=-l_b}^{l_b}
 S\left(
 \begin{array}{ccc}
  0&0&0 \\
  l_a&l_b&l_c\\
  m_a&m_b&m_c
 \end{array}
 \right)
 Y_{l_a,m_a}(\vec{\Omega}_1)
 Y_{l_bm_b}(\vec{\Omega}_2)Y_{l_cm_c}(\vec{\Omega}_3),
\end{align}
where we have used
$\delta^{(2)}(\vec{r}_1-\vec{r}_2)\delta^{(2)}(\vec{r}_1-\vec{r}_3)
=\delta^{(2)}(\vec{r}_3-\vec{r}_1)\delta^{(2)}(\vec{r}_3-\vec{r}_2)$.
The matrix element is
\begin{align}
 &(\bra{l_1,m_1}\otimes\bra{l_2,m_2}\otimes\bra{l_3,m_3})
 \hat{V}_3^{(s,t,u)}
 (\ket{l'_1,m'_1}\otimes\ket{l'_2,m'_2}\otimes\ket{l'_3,m'_3})\non
 =&\sum_{l_a,l_b=0}^\infty
 \sum_{l_c=|l_a-l_b|}^{l_a+l_b}
 \left[-l_a(l_a+1)\right]^{s}\left[-l_b(l_b+1)\right]^{t}
 \left[-l_c(l_c+1)\right]^{u}
 \sum_{m_a=-l_a}^{l_a}\sum_{m_b=-l_b}^{l_b}
 S\left(
 \begin{array}{ccc}
  0&0&0 \\
  l_a&l_b&l_c\\
  m_a&m_b&m_c
 \end{array}
 \right)
 \int d\vec{\Omega}_1d\vec{\Omega}_2d\vec{\Omega}_3\non
 &\qquad
 \left[
 Y_{Ql_3m_3}^*(\vec{\Omega}_3)
 Y_{Ql_2m_2}^*(\vec{\Omega}_2)
 Y_{Ql_1m_1}^*(\vec{\Omega}_1)
 \right]
 \left[
 Y_{l_a,m_a}(\vec{\Omega}_1)
 Y_{l_bm_b}(\vec{\Omega}_2)Y_{l_cm_c}(\vec{\Omega}_3)
 \right]
 \left[
 Y_{Ql'_1m'_1}(\vec{\Omega}_1)
 Y_{Ql'_2m'_2}(\vec{\Omega}_2)
 Y_{Ql'_3m'_3}(\vec{\Omega}_3)
 \right]\non
 =&(-1)^{3Q-m_1-m_2-m_3}
 \sum_{l_a,l_b=0}^\infty
 \sum_{l_c=|l_a-l_b|}^{l_a+l_b}
 \left[-l_a(l_a+1)\right]^{s}\left[-l_b(l_b+1)\right]^{t}
 \left[-l_c(l_c+1)\right]^{u}
 \sum_{m_a=-l_a}^{l_a}\sum_{m_b=-l_b}^{l_b}
 S\left(
 \begin{array}{ccc}
  0&0&0 \\
  l_a&l_b&l_c\\
  m_a&m_b&m_c
 \end{array}
 \right)\times\non
 &\qquad\qquad\qquad
 S\left(
 \begin{array}{ccc}
  -Q&Q&0 \\
  l_1&l'_1&l_a\\
  -m_1&m'_1&m_a
 \end{array}
 \right)
 S\left(
 \begin{array}{ccc}
  -Q&Q&0 \\
  l_2&l'_2&l_b\\
  -m_2&m'_2&m_b
 \end{array}
 \right)
 S\left(
 \begin{array}{ccc}
  -Q&Q&0 \\
  l_3&l'_3&l_c\\
  -m_3&m'_3&m_c
 \end{array}
 \right)\non
 =&
 (-1)^{3Q-m_1-m_2-m_3}
 \sum_{l_a=|l_1-l'_1|}^{l_1+l'_1}
 [-l_a(l_a+1)]^s
 S\left(
 \begin{array}{ccc}
  -Q&Q&0 \\
  l_1&l'_1&l_a\\
  -m_1&m'_1&m_a
 \end{array}
 \right)
 \sum_{l_b=|l_2-l'_2|}^{l_2+l'_2}
 [-l_b(l_b+1)]^t
 S\left(
 \begin{array}{ccc}
  -Q&Q&0 \\
  l_2&l'_2&l_b\\
  -m_2&m'_2&m_b
 \end{array}
 \right)\times\non
 &
 \sum_{l_c=\max\left[|l_3-l'_3|,|l_a-l_b|\right]}
 ^{\min\left[l_3+l'_3,l_a+l_b\right]}
 [-l_c(l_c+1)]^u
 S\left(
 \begin{array}{ccc}
  -Q&Q&0 \\
  l_3&l'_3&l_c\\
  -m_3&m'_3&m_c
 \end{array}
 \right)
 S\left(
 \begin{array}{ccc}
  0&0&0 \\
  l_a&l_b&l_c\\
  m_a&m_b&m_c
 \end{array}
 \right)
 \Biggr|_{m_a=m_1-m'_1,m_b=m_2-m'_2,m_c=-m_a-m_b}.
 \label{eq:123123SUM}
\end{align}
\end{widetext}
One gets Eq.~\eqref{eq:123123} by defining 
$m_1+m_2+m_3$ and $\left[-l(l+1)\right]^tS$ as $m_{123}$ and $S^{(t)}$, 
respectively. The calculation of the elements with large $2Q$ is a 
time-consuming task. The expression in Eq.~\eqref{eq:123123SUM}
should be employed rather than that in Eq.~\eqref{eq:123123} to reduce the number of times 
$S$ is computed in numerical calculations.

\bibliography{biblio_fqhe.bib}
\bibliographystyle{apsrev}

\end{document}